\documentclass{ws-mpla}
\usepackage[super]{cite}
\usepackage{graphicx}

\usepackage[hidelinks]{hyperref}
\hypersetup{
    colorlinks=false,
    linkcolor={red!50!black},
    citecolor={blue!50!black},
    urlcolor={blue!80!black}
} 
\newcommand{\be}{\begin{equation}}
\newcommand{\ee}{\end{equation}}
\newcommand{\bs}{\begin{split}} 
\newcommand{\bea}{\begin{eqnarray}}
\newcommand{\eea}{\end{eqnarray}}

\begin{document}

\markboth{M. Good, E. Linder \& F. Wilczek}
{Finite Thermal Particle Creation of Casimir Light}

\catchline{}{}{}{}{}

\title{FINITE THERMAL PARTICLE CREATION OF CASIMIR LIGHT}

\author{\footnotesize MICHAEL R. R. GOOD\footnote{
This proceedings contribution is based on a June 28, 2019 talk given by MRRG at the 4th Casimir Symposium in St. Petersburg, Russia. The author's present address is Nur-Sultan, Kazakhstan.}}
\address{Physics Department \& ECL, Nazarbayev University,\\ Nur-Sultan, 010000, Kazakhstan.\\ 
michael.good@nu.edu.kz}

\author{ ERIC V. LINDER,$^{1,2}$\, FRANK WILCZEK$^{3,4,5,6}$}
\address{$^{1}$Energetic Cosmos Laboratory, Nazarbayev University,\\ Nur-Sultan, 010000, Kazakhstan.\\
$^{2}$Berkeley Center for Cosmological Physics \& Berkeley Lab, University of California, \\Berkeley, California, 94720, USA.\\
$^{3}$Center for Theoretical Physics, Massachusetts Institute of Technology,\\
Cambridge, MA, 02139, USA.\\
$^{4}$T. D. Lee Institute \& Wilczek Quantum Center,
Shanghai Jiao Tong University, \\
Shanghai, 200240, China.\\
$^{5}$Department of Physics \& Origins Project, Arizona State University, \\
Tempe, AZ, 25287, USA.\\
$^{6}$Department of Physics, Stockholm University,\\
Stockholm, SE-106 91, Sweden.}


\maketitle

\pub{Received 31 July 2019}{Revised (Day Month Year)}

\begin{abstract}
A new solution for an analytic spectrum of particle creation by an accelerating mirror (dynamical Casimir effect) is given.  It is the first model to simultaneously radiate thermally and emit a finite number of particles. 

\end{abstract}


\section{Trajectory}
Evaporating black holes,\cite{Hawking:1974sw} acceleration radiation,\cite{Unruh:1976db}. and moving mirrors,\cite{Davies:1976hi} have no known particle production model that has global finite count $N$ \textit{and} thermal radiation spectra  $N_\omega$. We present such a spectrum via a new\cite{Good1} and simple moving mirror. 

 Particle production models that give an exact Planck spectrum of thermal radiation with a temperature $T = \kappa/2\pi$ (where $\kappa$ sets the scale -the surface gravity in the black hole context) are numerous, but all models with exact thermal radiation have total divergent particle production.\footnote{Large but finite particle emission in a Friedmann universe has been calculated.\cite{Grib}}  This infinity is no problem when  calculating the particle density,\cite{Grib} 
 but necessarily involves an infrared divergence in the spectrum resulting in infinite soft particles with zero frequency.\footnote{Soft particles signal a remnant \cite{Chen:2014jwq} in the moving mirror model.}  
 
The moving mirror model (see e.g. some recent important works \cite{Fulling:2018lez,harvest}) has the Carlitz-Willey trajectory\cite{Carlitz:1986nh}, which is an eternally thermal Planckian motion with a dynamic but simple acceleration \cite{paper2}, $\alpha(\tau) = c/\tau$, 
where 
$\tau$ is the proper time. The global particle count diverges because the mirror accelerates forever. 
The divergence is more widely associated with the assumption of asymptotically null trajectories \cite{Good:2013lca, spin}.  Naively, one might try to solve this by constructing asymptotically static trajectories, \cite{walkerdavies, purity,paper1,paper3} and they do in fact produce a finite amount of particles; however, the drawback is no Planck distribution!

Therefore it would be worthwhile to have a solution with finite thermal particles, with no light-like travel \cite{purity}. We construct this by building on the Schwarzschild mirror \cite{MG14one,MG14two,Good:2016LECOSPA,Good:2016MRB}, which has a one-to-one correspondence to the black hole, with trajectory $\kappa v = -e^{2 \kappa  x}$, where $v$ is advanced null time, $v = t +x$, and $(x,t)$ are lab space-time coordinates, $c=1$.  
This is a light-like trajectory that suffers from information loss,\footnote{Entanglement entropy plays an important role in particle creation models, and takes particularly simple forms in the moving mirror model. \cite{Chen:2017lum,horizonless,Holzhey:1994we,Hwang:2017yxp, paper3}} infinite particle production, and infinite energy production. However, at late times the spectra is exactly Planckian with temperature $ T = \kappa/2\pi$.  Reflecting this reflector, and introducing a new dimensionful parameter $g$, the new trajectory is 
\be gv = -\sinh (2 \kappa  x). \label{FTP}\ee
This mirror is solvable and the thermal character is maintained when $g\gg \kappa$. It changes the light-like trajectory of the Schwarzschild mirror to an asymptotically static worldline (time-like path), resulting in a finite total particle production.   
\section{Spectrum}
\begin{figure}[ht]
\centering
\includegraphics[width=3.2in]{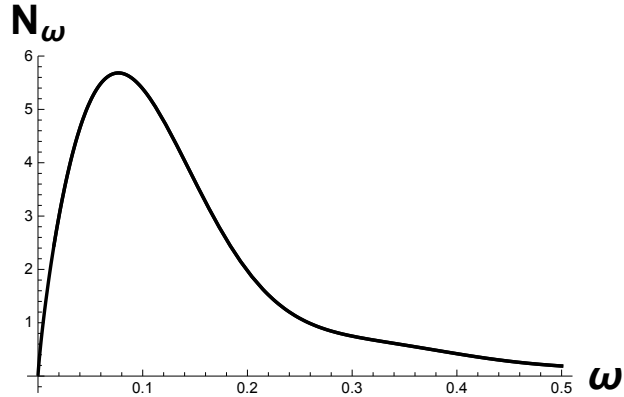} 
\caption{The spectrum, $N_\omega$, Eq.~(\ref{full}), detected at right null infinity, $I^+_R$, of the 
asymptotically static thermal mirror is plotted vs frequency $\omega$, in units of $\kappa=1$.  
The 1+1 dimensional thermal spectrum from this asymptotically static solution has no infrared divergence (no soft particles as $\omega\to0$) similar to a typical 3+1 dimensional blackbody curve; here $g> \kappa$ with $g/\kappa = 10^6$. 
\label{fig:Particle_Spectrum}} 
\end{figure}

Using the trajectory Eq.~(\ref{FTP}) in the beta Bogolubov coefficient integral,\cite{horizonless}
\be \beta_{\omega\omega'} =\frac{-1}{4\pi\sqrt{\omega\omega'}}\int_{+\infty}^{-\infty} dx\; e^{2 i \omega x - i\omega_p v(x)}(\omega_n v'(x) - 2\omega), \ee
results in a new beta with analytic expression \cite{Good1},
\be \beta_{\omega\omega'} = -\frac{\sqrt{\omega\omega'}}{\pi  \kappa  \omega_p }e^{-\frac{\pi  \omega }{2 \kappa }} K_{\frac{i \omega }{\kappa }}\left(\frac{\omega_p}{g}\right) ,\label{beta}\ee
where $K_n(z)$ is a modified Bessel function of the second kind, and $\omega_{p,n} \equiv \omega \pm \omega'$.  The beta coefficient is used for calculating the particle spectrum, $N_\omega$, detected at right null infinity surface  $I^{+}_R$ by integrating the modulus squared of Eq.~(\ref{beta}),  
\be N_\omega = \int_0^\infty  |\beta_{\omega\omega'}|^2 \; d \omega'\ . \label{N_w}\ee
The result for $N_\omega$ is analytic but complicated.\footnote{The hypergeometric result has a non-thermal form of energy flux when $g\sim \kappa$.} However, to leading order for large $g \gg \kappa =1$, the result has a temperature $T=1/2\pi$ with:
\be N_\omega = \frac{\bar{\Gamma}_\omega}{e^{ \omega/T}-1},\label{full}\ee
where $\bar{\Gamma}_\omega \equiv A/\pi + B/4$, $A \equiv \ln \left(2g/\omega\right)+\Re\left(H_{i \omega }\right)-\gamma -1$, $H_n$ is the Harmonic number, $\gamma = 0.577$ is Euler's constant, and  
\be B \equiv \frac{\left(\omega/2g\right)^{2 i \omega } \text{csch}(\pi  \omega )}{(2 \omega +i) \Gamma (1+i \omega)^2}+\frac{\left(\omega/2g\right)^{-2 i \omega } \text{csch}(\pi  \omega )}{(2 \omega -i) \Gamma (1-i \omega )^2}.\ee
Fig.~\ref{fig:Particle_Spectrum} illustrates the resulting spectrum $N_\omega$, Eq.~(\ref{full}), showing the desired absence of infinite soft particles in the infrared regime, $\omega\rightarrow 0$.

The thermal nature of the particle emission is evident from the Planck factor in Eq.~(\ref{full}), as $\bar{\Gamma} \rightarrow \ln(2g)/\pi = \textrm{constant}$, when $\omega\gg \kappa = 1$.  Wave packets can be used to resolve the thermal particle emission in frequency or time. \cite{Hawking:1974sw,Good:2013lca,Good:2016HUANG,signatures} 
Eq.~(\ref{full}) allows the total particle count to be obtained by 
a single numerical integration of $N_\omega$ over $\omega$,
\be
N = \int_0^\infty N_{\omega} \; d\omega\ , \label{totalparticles}
\ee
rather than a double 
numerical integration of the Bogolubov coefficients, $|\beta_{\omega\omega'}|^2$, over both $\omega$ and $\omega'$. Integration of Eq.~(\ref{totalparticles}) with ultra-relativistic speed ($g \gg \kappa$) gives the most particle production.  A finite large total emission of particles obtained here is similar to what is expected from an effervescent thermal black hole\cite{Good2}, leaving no trace of a remnant \cite{GTC,universe,MG15}.

\section*{Acknowledgments}

MG is funded from Kazakhstan MES Grant BR05236454 and by the ORAU Research Grant No. 090118FD5350. Thanks is given to the organizers, in particular Galina Klimchitskaya, Vladimir Mostepanenko, and the anonymous reviewers of the Proceedings of the 4th Casimir Symposium, St. Petersburg, Russia, 2019.


\end{document}